\documentclass[%
 reprint,
 amsmath,amssymb,
 aps,
]{revtex4-2}

\usepackage{graphicx}
\usepackage{dcolumn}
\usepackage{bm}

\begin{document}

\preprint{APS/123-QED}

\title{A clock with $8\times10^{-19}$ systematic uncertainty}
\author{Alexander Aeppli$^1$, Kyungtae Kim$^1$, William Warfield$^1$, Marianna S. Safronova$^2$, and Jun Ye$^1$}
 \email{Alexander.Aeppli@colorado.edu, Ye@jila.colorado.edu}
\affiliation{$^1$~JILA, National Institute of Standards and Technology and the University of Colorado, Boulder, Colorado 80309-0440, USA \\
and Department of Physics, University of Colorado, Boulder, Colorado 80309-0390, USA}
\affiliation{$^2$~Department of Physics and Astronomy, University of Delaware, 19716, USA}

\date{\today}

\begin{abstract}
We report an optical lattice clock with a total systematic uncertainty of $8.1 \times 10^{-19}$ in fractional frequency units, representing the lowest uncertainty of any clock to date. 
The clock relies on interrogating the ultra-narrow ${}^1S_0 \rightarrow {}^3P_0$ transition in a dilute ensemble of fermionic strontium atoms trapped in a vertically-oriented, shallow, one-dimensional optical lattice. 
Using imaging spectroscopy, we previously demonstrated record high atomic coherence time and measurement precision enabled by precise control of collisional shifts and the lattice light shift.
In this work, we revise the black body radiation shift correction by evaluating the $5s4d$ $^3D_1$ lifetime, necessitating precise characterization and control of many body effects in the $5s4d$ $^3D_1$ decay.
Lastly, we measure the second order Zeeman coefficient on the least magnetically sensitive clock transition.
All other systematic effects have uncertainties below $1 \times 10^{-19}$.
\end{abstract}

\maketitle

\textbf{\emph{Introduction.}} Measuring time is one of the most fundamental tasks in physics, with each advancement in timekeeping enabling new discoveries and technologies~\cite{NAP25613, ludlowOpticalAtomicClocks2015}.
Owing to the higher frequency of electronic transitions, the exceptional stability of optical metrology promises to revolutionize many disparate fields, from fundamental physics to navigation and geodesy. 
Over the past two decades, optical atomic clocks using neutral atoms or single ions have surpassed those based upon microwave transitions, setting records for both stability and accuracy~\cite{mcgrewAtomicClockPerformance2018,bothwellJILASrIOptical2019,brewerAlAccuracy}.
Optical lattice clocks (OLCs) achieve ultra-high stability by simultaneously interrogating many atoms tightly confined within a standing wave of light~\cite{takamoto_optical_2005,ludlow_2006,bloom_optical_2014}. 
Every gain in stability and accuracy opens new realms of exploration, such as placing bounds on dark matter~\cite{filzinger_ybdm_2023,boulderatomicclockopticalnetworkbaconcollaboration*FrequencyRatioMeasurements2021}, probing general relativity~\cite{takamoto_test_2020,zheng_lab-based_2023}, and will ultimately result in the redefinition of the SI second~\cite{Dimarcq_2024,Riehle2015506,Lodewyck_2019}.
\par Building upon two decades of optical lattice clock development, the JILA strontium 1D OLC utilizes a shallow lattice formed within an in-vacuum build up cavity first described in Ref.~\cite{bothwellResolvingGravitationalRedshift2022}.
The $^{87}$Sr ${}^1S_0 \rightarrow {}^3P_0$ clock transition is addressed with a laser stabilized to a cryogenic, single-crystal silicon, optical resonator~\cite{oelkerDemonstration10172019}.
We previously reported record levels of atomic coherence and self-synchronous stability~\cite{bothwellResolvingGravitationalRedshift2022}, cancellation of atomic interaction shifts~\cite{aeppliHamiltonianEngineeringSpinorbit2022}, and precise control of the lattice light shift~\cite{kimSr1ACStark}.
In this Letter, we report a complete systematic evaluation with a total uncertainty of $8.1\times 10^{-19}$ in fractional frequency units. 
Improved measurements of the second order Zeeman coefficient and the dynamic shift for black body radiation allow us to make significant strides in clock accuracy. 
\begin{figure*}[th!]
    \includegraphics[width=17.78cm]{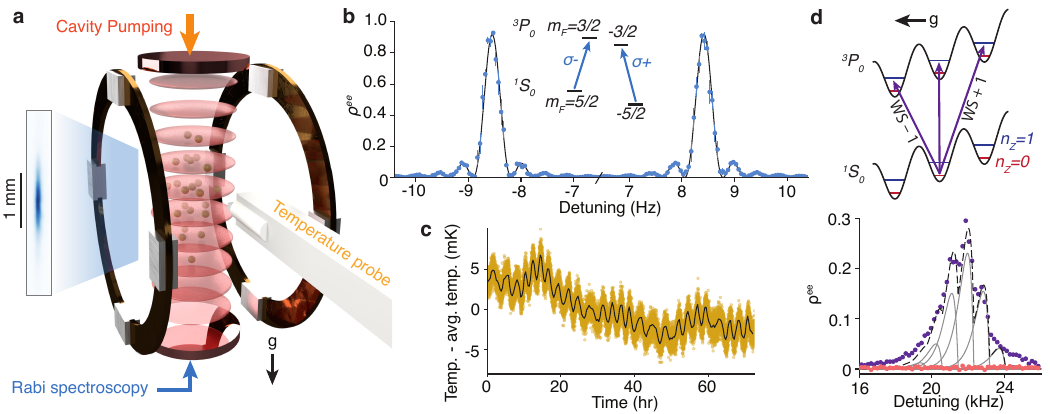}
    \caption{
        Overview of the 1D optical lattice clock.
        (a)  Schematic of the system highlights key aspects.
        Atoms are trapped in a 1D optical lattice formed within an in-vacuum buildup cavity oriented along the direction of gravity $g$.
        We readout the state of the atoms by imaging with a $6$~$\mu$m resolution.
        Rabi spectroscopy of the clock transition is performed along the tightly confined direction to remain within the resolved sideband regime. 
        The two circular rings are quadrant electrodes for applying an electric field in any direction.
        A translatable temperature probe with two sensors measures the temperature.
        (b) $2.43$~s Rabi spectroscopy of the two operational clock transitions. 
        We prepare atoms in the $^1S_0,~m_F = \pm 5/2$ states, drive the least magnetically sensitive clock transition to $^3P_0,~m_F = \pm 3/2$, and measure the excitation fraction $\rho^{ee}$.
        Blue points are an average of $5$ line scans with error bars given as the standard error. 
        The black line is a Rabi lineshape fit. 
        (c) Temperature measured at the atom location over 3 days. 
        The gold points are an average of the two sensors measured every 10 seconds, and the black line is a 20 minute average. 
        (d) Axial blue sideband (BSB) structure. 
        The tilted, shallow lattice creates a set of site-changing Wannier-Stark (WS) transitions where the $^3P_0$ state occupies a different lattice site than the $^1S_0$ state.
        At our operational depth, only two motional states along the tight confinement direction exist, $n_Z = 0$ and $1$.
        In the lower panel, purple points show the $n_Z = 0 \rightarrow 1$ transition.
        The structure of this spectrum is well captured by a model that incorporates the WS structure, the axial structure, and the radial temperature.
        Each gray line illustrates the BSB for each WS transition, with the sum of these in dashed black consistent with the data.
        In our standard clock sequence, the lattice depth is briefly reduced to $3$~$E_r$ before readout.
        The pink points near $0$ demonstrate that this approach effectively eliminates all but the ground band population.}
    \label{fig:1}
\end{figure*}
\par \textbf{\emph{High Accuracy Operation.}} To properly characterize and control systematic shifts, high measurement precision together with repeatable and reliable operation are key.
Critical environmental control begins by stabilizing the air flow around the vacuum chamber to $20$~$^{\circ}$C with $100$ mK peak-to-peak stability.
Each viewport flange on the vacuum chamber is temperature stabilized via separate liquid loops with better than $20$ mK stability. 
We directly measure the radiation temperature $T$ and its stability at the atom's location by translating in a pair of calibrated in-vacuum temperature sensors [Fig.~\ref{fig:1}(a)].
\par An effusive oven generates a collimated beam of $^{87}$Sr that is slowed and cooled using the broad $^1S_0 \rightarrow\,^1P_1$ transition at $461$ nm.
The oven does not have direct line-of-sight to the main vacuum chamber, and we measure no temperature coupling to the oven at the atoms.
This beam loads a magneto-optical trap (MOT) operating on this same transition.
Further cooling on the $^1S_0 \rightarrow\,^3P_1$ transition at $689$~nm reduces the temperature to a few $\mu$K~\cite{makotoSrMOT}.
This cooling light is stabilized to the same silicon resonator as the clock laser with Hz-level drift per day, making cooling robust over many months.
As shown in Fig.~\ref{fig:1}(a), a magic wavelength optical lattice at $813$~nm is formed within an in-vacuum buildup cavity oriented along gravity~\cite{bothwellResolvingGravitationalRedshift2022}.
Atoms are loaded from the MOT into the lattice at a depth of $300$ lattice photon recoil energies ($E_{r}$).
Doppler cooling on the $^1S_0 \rightarrow {}^3P_1$ $F=11/2$ transition reduces the radial temperature to $\sim 700$ nK, and resolved motional sideband cooling along the tightly confined direction $\hat{Z}$ reduces the expected quantum state to $\langle n_Z \rangle < 0.05$.
During Doppler and sideband cooling, we polarize the the atomic sample in one of the $m_F = \pm 9/2$ stretched states.
\par To reduce both the lattice light shift and the density shift, it is generally optimal to operate at a shallow lattice depth. 
Further, we identified a ``magic lattice depth'' near $15$ $E_r$ where on-site interactions are cancelled by off-site interactions, leading to a net-zero density shift~\cite{aeppliHamiltonianEngineeringSpinorbit2022}.
Adiabatically ramping the lattice from the loading depth to $15$ $E_r$ reduces the radial temperature to $\sim 120$ nK.
The standard motional sideband model~\cite{blattRabiSpectroscopyExcitation2009} is no longer reliable at these depths. 
The blue sideband, corresponding with adding one motional quanta along $\hat{Z}$, splits due to transitions to neighboring lattice sites as shown in Fig. \ref{fig:1}(d). 
Thus it is necessary to use the lattice depth calibration technique introduced in Ref.~\cite{kimSr1ACStark}.
\par To reduce the Zeeman effect sensitivity, we use the least magnetically sensitive $|^1S_0 ~ m_F = \pm 5/2 \rangle \rightarrow | ^3P_0 ~ m_F = \pm 3/2 \rangle$ clock transition as illustrated in Fig.~\ref{fig:1}(b). 
Beginning with atoms in one of the stretched states, we apply two clock transfer pulses to prepare atoms in the $|^1S_0 ~ m_F = \pm 5/2 \rangle$ spectroscopy states with about $96 \%$ spin purity. 
\par Between sample preparation and readout, the dead time is $1$ s.
Based upon a noise model of the clock laser, Dick noise is minimized with a $2.43$~s Rabi interrogation time~\cite{oelkerDemonstration10172019}. 
A digital servo with two integrators tracks the atomic transition by alternating spin states and sides of the Rabi lineshape.
We expect a single clock stability of $5 \times 10^{-17}/\sqrt{\tau}$ for averaging time $\tau$ in seconds.
\par As shown in Fig.~\ref{fig:1}(d), we briefly reduce the lattice depth to the single band regime ($\sim 3$ $E_r$) before readout to ensure that the only measured atoms are in the $n_Z = 0$ state. 
High resolution imaging to readout the clock excitation resolves spatial frequency variation (Fig.~\ref{fig:1}(a)), allowing for real-time density shift corrections.
We operate with approximately $4 \times 10^4$ atoms, leading to a quantum projection noise of $< 3 \times 10^{-18}/\sqrt{\tau}$, which is near the self-synchronous comparison performance reported in Ref.~\cite{bothwellResolvingGravitationalRedshift2022}.
\par \textbf{\emph{Black body radiation shift.}} The largest systematic shift in room temperature Sr clocks arises from the black body radiation (BBR) environment. 
The total differential BBR shift $\Delta \nu_{\text{BBR}}$ is the sum of a static component $\nu_{\text{stat}}$ that scales as $T^4$, and a dynamic component $\nu_{\text{dyn}}$ that scales with higher powers of $T$.
Thus, for accurate operation we need to determine $T$ and the atomic response with high precision.
\par \emph{Radiation temperature.} To ensure a fully thermal environment and measure the radiant temperature at the atoms, we follow a similar technique as in Refs.~\cite{nicholsonSystematicEvaluationAtomic2015,bothwellJILASrIOptical2019}. 
Two calibrated thin film platinum resistance sensors are mounted to an in-vacuum translation arm, which is inserted into the middle of the vacuum chamber.
During clock spectroscopy the probe is retracted $30$~cm into an auxiliary vacuum chamber. 
We observe a sub-mK temperature flicker floor at short time scales, $\sim 2$~mK peak oscillations on the hour time scale, and drift of less than a few mK per day, as shown in Fig.~\ref{fig:1}(d). 
These few-hour temperature fluctuations are from coupling to room temperature and building process chilled water and can likely be improved by further system isolation.
At 12 hours the Allan deviation of the temperature is $1.4$~mK, which we treat as the operational stability.
The total temperature uncertainty is $4.1$~mK~\cite{SeeSupplementalMaterial}.
\par
\begin{figure}[t!]
    \includegraphics[width=8.6cm]{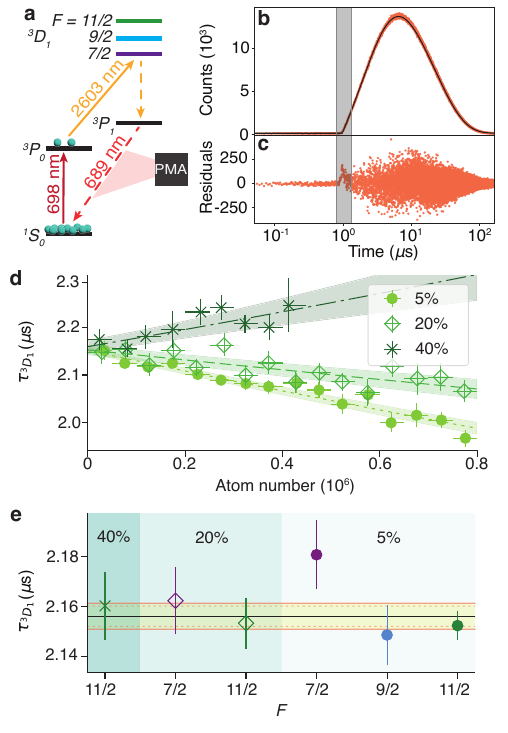}
    \caption{
    $5s4d~{}^3D_1$ lifetime measurement results. 
    (a) Atoms are prepared in the $^1S_0$ ground state, a fraction are pumped to $^3P_0$ and then excited to one of the $^3D_1$ hyperfine levels $F$. 
    Some of these atoms decay through $^3P_1$ back to $^1S_0$, where we collect and time resolve the photons on a hybrid photomultiplier assembly (PMA).
    (b) Fit of Eq. \eqref{eq:doubleExp} in black to all the collected data in orange.
    We exclude $500$~ns window around the $2.6$~$\mu$m excitation pulse, as shown in gray
    (c) Residuals of this fit. 
    (d) The measured $^3D_1$ lifetime $\tau_{^3D_1}$ as a function of total atom number for $^3D_1 ~ F=11/2$.
    The lines are density dependent lifetime fits with the shaded area showing the fit uncertainty. 
    Varying the portion of the atoms in $^3P_0$, shown as a percent of the total population, the density dependence of $\tau_{^3D_1}$ changes.
    (e) Results of the six different data sets. 
    Each point represents the zero density lifetime for different hyperfine levels $F$ and $^3P_0$ population fractions, labeled above.
    The black line is the weighted average, the dashed lines show the statistical uncertainty, and the solid orange lines show the combined statistical and systematic uncertainty.}
    \label{fig:3d1}
\end{figure}

 \emph{$\nu_{\text{dyn}}$ evaluation.} Accuracy in previous generations of room temperature Sr OLCs has been limited by the uncertainty in $\nu_{\text{dyn}}$, which is directly tied to the $5s4d$ $^3D_1$ lifetime~\cite{safronova_bbr_2013}.
As in Refs.~\cite{nicholsonSystematicEvaluationAtomic2015,beloyDynamic}, we prepare a sample of Sr atoms in $^3P_0$ before a $2.6$ $\mu$m laser pulse excites a portion of the sample to the $^3D_1$ state. 
Some excited atoms decay to $^3P_1$ and then to the $^1S_0$ ground state, releasing a $689$ nm photon in the process, as shown in Fig.~\ref{fig:3d1}(a).
We collect this fluorescence with a cooled hybrid photomultiplier assembly (PMA) and time tag the incident photons with $5$~ns resolution.
In the single particle regime, the photon rate $y$ at time $t$ is well characterized by a cascaded double exponential process,
\begin{equation}
\label{eq:doubleExp}
    y(t) = A \times \Theta(t-t_0) \left( e^{-(t-t_0)/\tau_{^3D_1}}-e^{-(t-t_0)/\tau_{^3P_1}}\right) + y_0,
\end{equation}
where $A$ is the flux amplitude, $\Theta$ is a Heaviside function for instantaneous excitation at time $t_0$, $\tau_{^3D_1}$ and $\tau_{^3P_1}$ are the $^3D_1$ and $^3P_1$ lifetimes respectively, and $y_0$ is an offset due to background counts. 
In Fig.~\ref{fig:3d1}(b) we plot all collected photon counts and fit with Eq.~\eqref{eq:doubleExp}.
Since Eq.~\eqref{eq:doubleExp} assumes instantaneous excitation of atoms to $^3D_1$, we do not fit data within a $500$~ns window about the excitation pulse, indicated by the gray exclusion area~\cite{nicholsonSystematicEvaluationAtomic2015}. 
\par At high densities in the $^3P_0$ state, we notice a modification to the exponential decay process.
Spontaneous emission from one atom can affect the behavior of another atom, leading to effective dipole-dipole interactions and giving rise to effects like superradiance or radiation trapping. 
The interplay of these effects in this cascaded, multi-state decay is hard to simulate theoretically, and we do not have a complete model for extracting single particle lifetimes when such effects are present.
\par To use the model in Eq.~\eqref{eq:doubleExp} to determine the lifetime, it is vital to keep the population in $^3P_0$ low as it is the primary state that contributes to collective effects in the $^3D_1$ decay process.
However, reducing atom number adversely affects averaging time. 
Instead, we load a large number of atoms in $^1S_0$ and promote a small portion of the atoms to $^3P_0$. 
We then excite these atoms to $^3D_1$ with a $100$ ns laser pulse.
Since most of the atoms decay back to $^3P_0$, we repeat this process $15$ times before again exciting a portion of the $^1S_0$ atoms to $^3P_0$. 
After $10$ clock pulses and a total of $150$ $^3D_1$ decay cycles, we Doppler cool the remaining sample. 
We repeat this excitation and cooling sequence $5$ times before reloading a sample into the lattice. 
In sum, for each MOT sequence we collect photons from $750$ decay cycles. 
On average we capture less than one photon from the sample per decay cycle, so pile-up effects are effectively eliminated.
\par Although the population that contributes to non-linear collective effects is significantly reduced, weaker collective effects are still present.
To understand how density affects the measured lifetime, we measure the total atom number at the beginning and end of the sequence and assign an atom number to each decay event. 
\par Over the course of the measurement campaign, we collect $8\times10^7$ photons with a total atom number ranging from $10^3$ to $10^6$.
We collect data with different proportions of atoms in $^3P_0$ and iterate over all three hyperfine levels in $^3D_1$, for a total of six separate data sets~\cite{SeeSupplementalMaterial}.
We divide each data set by atom number into bins with widths of $5 \times 10^4$ atoms and fit Eq.~\eqref{eq:doubleExp} to each bin. 
As in Ref.~\cite{beloyDynamic}, we fit the lifetime density dependence $\tau (n) = \tau_0/(1+cn)$, where $n$ is the camera measured atom number and is proportional to density, $\tau_0$ is the single atom lifetime, and $c$ is the density dependence coefficient.
At very high density, we notice that the data deviates from the linear model, so we choose to exclude data above $8\times 10^5$ atoms in these fits. 
This choice does not change the final reported value. 
With small population in the $^3P_0$ state, the measured lifetime is shorter at higher density. However, by initially placing $40\%$ of the population in $^3P_0$ the trend is reversed and higher density results in a longer observed lifetime, as shown for the $F=11/2$ data in Fig.~\ref{fig:3d1}(d). 
The single atom lifetime for these six data sets is plotted in Fig.~\ref{fig:3d1}(e).
\par A number of other systematics can modify the measured lifetime. 
By applying a magnetic field ($B$) of $\sim 1$~G, we observe Zeeman beats due to interference between the emitted photons. 
We limit this effect by operating at zero field and periodically measure and correct the background $B$ $< 2$~mG.
A similar effect is caused by lattice light shifts splitting magnetic sublevels. 
To reduce this systematic, we measure the decay in a $10$~$E_r$ lattice with the PMA oriented within a few degrees of the lattice light polarization.
Modeling these effects we assign an uncertainty of $3$~ns due to potential Zeeman beats. Using a weighted average of the six experimental conditions, we report $\tau_{^3D_1} = 2.156 \pm 0.005$~$\mu$s.
This is plotted as the solid black line in Fig.~\ref{fig:3d1}(e), with the statistical uncertainty shown as the dashed orange lines and the total uncertainty as the solid orange lines.
\par This precisely determined $^3D_1$ lifetime allows us to reevaluate $\nu_{dyn}$ based on the technique described in Ref.~\cite{lisdat_bbr_2021}.
Using known atomic properties including measured transition strengths, magic wavelengths, and static polarizability, we determine $\nu_{dyn} = -153.06(33)$~mHz at 300~K~\cite{SeeSupplementalMaterial}.
The final BBR-related frequency shift combining both the static~\cite{middelmann_static_2012} and dynamic effects at the operational temperature of $20.132(4)$~$^{\circ}$C is $(-48417.2 \pm 7.3)\times 10^{-19}$.
\par \textbf{\emph{Lattice light shift.}} In previous work~\cite{kimSr1ACStark}, we have demonstrated the ability to control the lattice light shift to a few parts in $10^{-19}$. 
Due to the differential sensitivity to the lattice light shift of the motional states along the tightly confined direction, care must be taken to ensure repeatable cooling of the sample in the lattice. 
The last stage of cooling is robust and stable, and we take a further step to reduce sample uncertainty by ramping the lattice to $3$~$E_r$ before readout, ensuring that only the lowest band population is measured as shown in Fig.~\ref{fig:1}(d). 
With identical atomic coefficients as in Ref.~\cite{kimSr1ACStark}, a lattice depth of $15.06(17)$~$E_r$, and a $10.5$~MHz lattice detuning from the measured operational magic frequency~\cite{ushijimaOperationalMagicIntensity2018}, the total light shift uncertainty is $3.2 \times 10^{-19}$.
\par \textbf{\emph{DC Stark Shift.}} Stray electric fields can shift the clock transition frequency~\cite{Lodewyck_dcstark_2012}.
To limit the possibility of patch charges on the mirror surfaces causing these shifts, Faraday shields surround the mirrors and provide passive field attenuation.
A pair of in-vacuum quadrant electrodes can apply electric fields in any direction across the atomic sample, shown as copper rings in Fig.~\ref{fig:1}(a). 
Alternating high and low fields, we precisely measure the residual DC Stark shift.
The shift is below $10^{-21}$ along the cavity direction. 
The dominant source of residual field is along the imaging axis–-likely due to patch charges on the large vacuum window nearby the atoms.
The total residual DC Stark shift is $-9.8 \pm 0.7 \times 10^{-20}$.
\par \textbf{\emph{Zeeman shifts. }} Due to the differential Land\'e-$g$ factor~\cite{boydNuclearSpinEffects2007} between the $^1S_0$ and $^3P_0$ states, we are sensitive to Zeeman shifts on the clock transition. 
Probing opposite spin states and taking the frequency average, we broadly reject this systematic.
Yet there is still sensitivity to magnetic field fluctuations at and below the experiment cycle frequency.
By using the $|^1S_0 ~ m_F = \pm 5/2 \rangle \rightarrow | ^3P_0 ~ m_F = \pm 3/2 \rangle$ transition, we substantially reduce coupling to the magnetic environment, however even small field drifts may cause frequency shifts.
\par We use the $26$ times more magnetically sensitive $|^1S_0 ~ m_F = - 5/2 \rangle \rightarrow | ^3P_0 ~ m_F = - 7/2 \rangle$ transition to characterize this effect. 
Measuring this transition with the same duty cycle as in standard operation, the frequency difference between alternating cycles gives an upper bound on the first order Zeeman shift. 
We find a flicker floor of $0.78$~mHz, leading to a total Zeeman shift uncertainty on the operational transition of $7 \times 10^{-20}$. 
\par Operation on the $|^1S_0 ~ m_F = \pm 5/2 \rangle \rightarrow | ^3P_0 ~ m_F = \pm 3/2 \rangle$ transition requires reevaluation of the second order Zeeman coefficient for our desired accuracy goal. 
This shift $\Delta \nu_{Z2}$ goes as
\begin{equation}
\label{eq:Z2}
    \Delta \nu_{Z2} = \xi_{\sigma\,m_F=5/2} \left(\Delta_{\text{meas}} - \Delta_{\text{vec}} \right)^2,
\end{equation}
where $\xi_{\sigma \, m_F = 5/2}$ is the second order Zeeman coefficient, $\Delta_{\text{meas}}$ is the measured frequency difference between the operational transitions, and $\Delta_{\text{vec}}$ is the splitting due to the lattice vector shift. 
\par
\begin{figure}[t!]
    \includegraphics[width=\columnwidth]{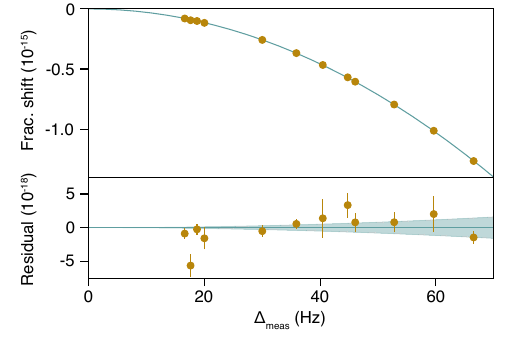}
    \caption{
        Second order Zeeman shift coefficient measurement. 
        We vary the applied magnetic field and measure the splitting ($\Delta_{\text{meas}}$) and the frequency shift (see also Fig.~\ref{fig:1} (b)).
        We fit the data with Eq. \eqref{eq:Z2} and plot this fit in green. 
        The lower panel shows the fit residuals with the shaded green region representing the fit uncertainty.
        }
    \label{fig:3}
\end{figure}

 To determine $\xi_{\sigma \, m_F = 5/2}$ precisely, we vary the applied bias field from $0.3$ to $1.5$~G and measure the resultant frequency shift in an interleaved manner, as shown in Fig.~\ref{fig:3}.
$\Delta_{\text{vec}}$ is measured independently by modulating the lattice depth.
We find a $\xi_{\sigma m_F = 5/2} = (-0.12263\pm 0.00014)$~mHz/Hz$^2$.
At the operational field near $380$~mG, the second order Zeeman shift $\Delta \nu_{Z2} = (-85.51\pm 0.10)\times10^{-18}$. 
\par \textbf{\emph{Tunneling Shift. }} At shallow depths, superposition of states in neighboring sites can cause frequency shifts.  
First identified in Ref.~\cite{lemondeOpticalLatticeClock2005}, the maximum possible frequency shift due to this effect goes as $\Omega_0 \Omega_1/ \Delta_g$, where $\Omega_0$ and $\Omega_1$ are the Rabi frequencies of the carrier and first Wannier-Stark sideband and $\Delta_g$ is the frequency difference between neighboring lattice sites.
At the operational depth of $15$~$E_r$, the off-site Rabi frequency is appreciably large, leading to a maximum shift of  $\sim 2 \times 10^{-19}$.
While the coherent superposition of neighboring states is likely small, it is difficult to directly measure and control this effect.
Instead, we opt to use a Rabi pulse time that is a half integer multiple of the tunneling shift oscillation period~\cite{lemondeOpticalLatticeClock2005}.
We measure the splitting between neighboring lattice sites to be $867.7461 \pm 0.0004$~Hz.
With a pulse time of $2.4298583$~s and a conservative timing uncertainty of $1$~$\mu$s, the maximum tunneling shift is $2 \times 10^{-21}$.
\par \textbf{\emph{Density Shift.}} Although strong collisional shifts are suppressed by the fermionic nature of $^{87}$Sr, $p$-wave interactions lead to a systematic density shift~\cite{swallows_collisions_2011,Martin2013}.
Previous Sr OLCs required separate evaluation of collisional shifts, leading to drifting systematics as sample preparation varied over time~\cite{bothwellJILASrIOptical2019}.
With imaging synchronously measuring different densities and frequency shifts throughout the sample, we perform real-time density shift corrections. 
As reported in Ref.~\cite{aeppliHamiltonianEngineeringSpinorbit2022}, operating at a ``magic lattice depth'' near $15$~$E_r$, on-site $p$-wave and off-site $s$-wave interactions cancel each other, substantially reducing the density shift even with a large atomic sample.
For a single characteristic run $\sim 300$~minutes, the correction is  $(-1.1 \pm 0.9) \times 10^{-19}$.
\par \textbf{\emph{Other Systematic Shifts.}} Collisions between trapped strontium atoms and background gas result in a systematic frequency shift~\cite{gibble_bgshift}.
In this system, the background gas is dominated by hydrogen molecules~\cite{SeeSupplementalMaterial}. 
As demonstrated in~\cite{alves_bgshift}, the shift is inversely proportional to the vacuum lifetime.
Using this coefficient and a measured vacuum lifetime of $63.6 \pm 2.5$~s, we calculate a background gas shift of $(-4.7 \pm 0.5)\times 10^{-19}$.
\par Line pulling occurs if population in other magnetic sublevels is off-resonantly driven, distorting the carrier lineshape.
With a low intensity $2.43$~s clock pulse, other transitions are highly suppressed.
With $96\%$ of the sample in the desired magnetic sublevel, we estimate for the worst case scenario a line pulling shift of $< 10^{-21}$.
\par Similarly, a low intensity Rabi drive significantly reduces the light shift from the clock laser. 
Using the coefficient measured in Ref.~\cite{Xu_probeStark}, and accounting for the increased intensity due to both light polarizations, we estimate the probe AC stark shift to be $-4 \times 10^{-22}$, which we treat as the uncertainty.
\par Thermal transients in the acousto-optic modulator (AOM) due to switching may lead to an uncorrected Doppler shift. 
As in~\cite{bothwellJILASrIOptical2019}, we path length stabilize the same AOM order that drives the atomic transition.
The AOM is ramped onto resonance after these thermal transients have settled, leading to an estimate probe chirp shift $< 10^{-21}$. 
\def\arraystretch{1.02}
\begin{table}[t]
\caption{\label{tab:ErrorBudget}%
Fractional frequency shifts and uncertainties for the JILA 1D Sr optical lattice clock.}
\begin{tabular}{lccr}
\hline \hline
         Shift Name &  Shift ($10^{-19}$) &  Uncertainty ($10^{-19}$)  \\
\hline

BBR &              -48417.2 &                               7.3 \\
      Lattice Light &                  -0.1 &                               3.2 \\
Second Order Zeeman &                -855.1 &                               1.0 \\
            Density &                  -1.1 &                               0.9 \\
         First order Zeeman &                   0.0 &                               0.7 \\
         Background Gas &                  -4.7 &                               0.5 \\
           DC Stark &                  -1.0 &                               0.1 \\
          Tunneling &                   0.0 &                               $<$0.1 \\
         Minor Shifts &                   0.0 &                               $<$0.1 \\ \hline
        Total Shift &              -49279.2 &                               8.1 \\

\hline \hline
\end{tabular}
\end{table}
\par \textbf{\emph{Summary.}} Through precise atomic and environmental control, we have realized a strontium optical lattice clock with a total systematic accuracy of $8.1 \times 10^{-19}$ as reported in table~\ref{tab:ErrorBudget}.
This represents greater than a factor of $2$ improvement in systematic accuracy over the previously most accurate strontium optical lattice clock~\cite{bothwellJILASrIOptical2019}, and it sets the accuracy benchmark of all optical clocks reported to date. Black body radiation stands out as the most significant source of uncertainty, and future cryogenic operation should reduce uncertainty to the low $10^{-19}$ level~\cite{ushijimaCryogenicOpticalLattice2015}.
\par \textbf{\emph{Acknowledgements.}} 
We thank S. Porsev for accurate atomic structure calculations, D. Wellnitz and A. M. Rey for useful discussions on collective radiation effects, and C. Lisdat and U. Sterr for discussions about the dynamic BBR correction.
We acknowledge contributions and discussions from the JILA Sr team, T. Bothwell, C. Kennedy, E. Oelker, J. Robinson, W. Tew, and A. Ellzey.
We thank A. Chu and N. Darkwah Oppong for close reading of this manuscript.
This research was supported in part through the use of University of Delaware HPC Caviness and DARWIN computing systems.
Funding support is provided by NSF QLCI OMA-2016244, V. Bush Fellowship, the DOE Quantum Systems Accelerator, NIST, and NSF PHY-2317149. K. K. is supported in part by the Quantum Information Research Support Center funded by NRF Korea, MSIT- No.2021M3H3A103657313. 


\bibliography{bibliography}

\end{document}


\newcommand{\Sr}{${}^{87}$Sr}
\newcommand{\Er}{$E_{r}$}
\renewcommand{\thefigure}{S\arabic{figure}}\makeatother
\renewcommand{\thetable}{S\arabic{table}}\makeatother


\title{Supplementary Material: A clock with $8\times10^{-19}$ systematic uncertainty}
\author{Alexander Aeppli$^1$, Kyungtae Kim$^1$, William Warfield$^1$, Marianna S. Safronova$^2$, and Jun Ye$^1$}
 \email{Alexander.Aeppli@colorado.edu, Ye@jila.colorado.edu}
\affiliation{$^1$~JILA, National Institute of Standards and Technology and the University of Colorado, Boulder, Colorado 80309-0440, USA \\
and Department of Physics, University of Colorado, Boulder, Colorado 80309-0390, USA}
\affiliation{$^2$~Department of Physics and Astronomy, University of Delaware, 19716, USA} 
\maketitle

\section{$^3D_1$ Lifetime}
To perform the $^3D_1$ lifetime measurement, we begin by preparing a $\sim 200$~nK ground band sample of atoms in a $10$~$E_r$ lattice. 
In contrast to our standard cooling sequence, the polarization light is turned off, and we verify an even spin mixture is obtained. 
A brief clock pulse excites $5\%$, $20\%$, or $40\%$ of this sample to the $^3P_0$ excited clock state. 
Then, a $100$~ns pulse generated by switching a $60$~MHz acousto-optic modulator excites atoms to the $^3D_1$ state.
The three hyperfine levels are well-resolved, and we tune the laser current to address each individually.
As we operate with a zeroed magnetic field, the direction of quantization is established by the lattice polarization, which is parallel to the incident direction of the $2.6$~$\mu$m light. 
The $2.6$~$\mu$m polarization is linear and along the lattice direction.
We collect the $^3P_1$ decay photons using a Picoquant cooled hybrid photomulitplier assembly (PMA hybrid-40) which has a quantum efficiency of $45\%$ at $689$~nm.
This detector is oriented $4 \pm 1.5^{\circ}$ from the excitation light direction. 
Two Semrock Brightline bandpass interference filters remove most of background light, and the background count rate is near the specified detector dark count rate of $80$~counts$/$s.
Incident photons are time tagged using a Picoquant TimeHarp 260 using $5$~ns timing resolution in its multi-channel scaler mode. 
\par For each decay event, we assign an atom number based upon blue fluorescence imaging done after the $750$ decay sequences and calibrated using the atomic quantum projection noise. 
Counts are sorted into bins by atom number with widths of $50$ thousand atoms, such that is there is one bin for atom numbers from $0-50,000$, one for $50,000-100,000$, and so on. 
Each atom number bin is fit with the double exponential decay equation using maximum likelihood estimation.
We found this fitting approach to be critical, as it properly accounts for the Poisson statistics governing time steps with low or no counts.
We fit the resultant lifetimes as a function of atom number using orthogonal distance regression to properly treat the atom number spread in each density bin. 
\par
\begin{figure}[th!]
    \includegraphics[width=10 cm]{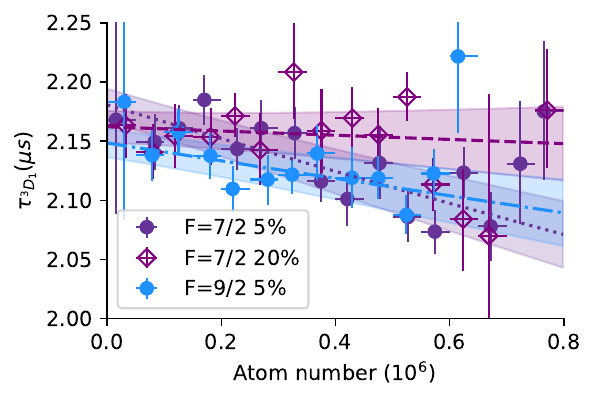}
    \caption{$^3D_1$ lifetime density dependence for the three data sets not presented in the main text.
    The $F$ levels are the $^3D_1$ hyperfine states and the percent corresponds with the initial population in $^3P_0$.
    The points are data, the lines are fits, and the shaded regions correspond with the fit uncertainty.
}
    \label{fig:3d1_slopes}
\end{figure}

 In Fig.~\ref{fig:3d1_slopes} we present the density dependence of the $^3D_1$ lifetime for two other hyperfine levels, $F=7/2$ and $F=9/2$, that were omitted from the main text. 
The points are lifetimes fit for each atom number bin, the lines are density dependent lifetime fits to the points, and the shaded area represents the fit uncertainty. 
As with the $F=11/2$ decay data, the $F=7/2$ data shows a shallower slope with an increased population ratio in the $^3P_0$ state.
\par Our result here notably differs from the measurement in Ref.~\cite{nicholsonSystematicEvaluationAtomic2015}, which found $\tau_{^3D_1} = 2.18(1)$~$\mu$s.
This discrepancy may be explained by the different regimes of these measurements. 
In Ref.~\cite{nicholsonSystematicEvaluationAtomic2015}, the experiment was performed with the population entirely in the $^3P_0$ state, leading to a higher effective optical depth (OD) on the $^3D_1 \rightarrow ^3P_0$ decay along the collection axis.
With a similar OD we observe longer lifetimes.
For our measurements, we estimate an OD on the $^3D_1$ decay with $100\%$ of the atoms in $^3P_0$ to be roughly $12$ per million atoms.
In Ref.~\cite{nicholsonSystematicEvaluationAtomic2015}, we estimate an OD of $1000$ per million atoms. 

\par
\begin{figure}[h!]
    \includegraphics[width=10 cm]{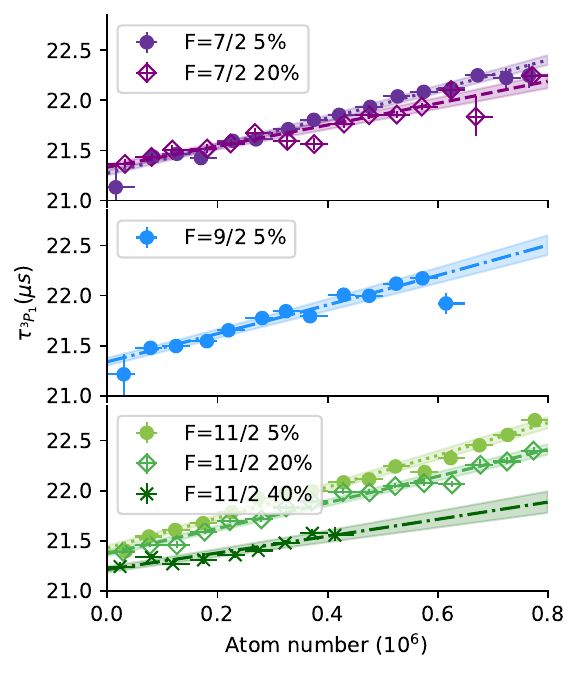}
    \includegraphics[width=10 cm]{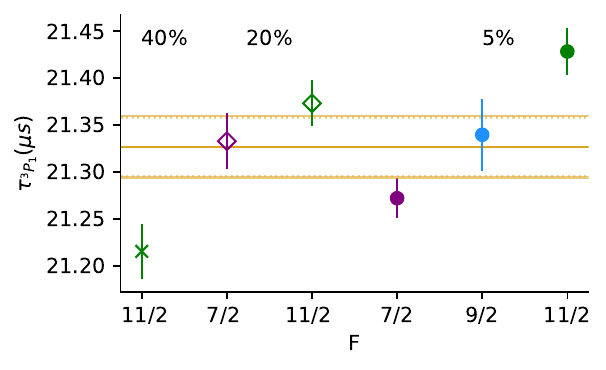}
    \caption{The $^3P_1$ lifetime results.
    The top panel shows the density dependent lifetime $\tau_{^3P_1}$ for the six different data sets. 
    The bottom panel is the single atom extracted $\tau_{^3P_1}$ organized by hyperfine level $F$ and the population fraction in $^3P_0$. 
    The darker orange line is the average lifetime, the dashed orange lines represent the statistical uncertainty, and the lighter orange lines are the combined statistical and systematic uncertainty.
}
    \label{fig:3p1 results}
\end{figure}

 The $^3P_1$ lifetime measurements extracted from the same data are presented in Fig.~\ref{fig:3p1 results}.
The top panel shows the density dependence of $\tau_{^3P_1}$ for all six data sets, the lines are fits, and the shaded regions show the fit uncertainty. 
As expected, the fraction of atoms in $^3P_0$ plays a smaller role in the density dependence of $\tau_{^3P_1}$, as this modification to the lifetime is broadly due to radiation trapping on the $^3P_1 \rightarrow\,^1S_0$ decay. 
With a larger portion of atoms in the $^3P_0$ state, fewer atoms participate in radiation trapping on this decay, leading to a shallower slope.
The extracted single atom lifetimes shown in the lower panel are more scattered than $^3D_1$.
The solid orange line represents the weighted average of all six data sets, the dashed lines show the statistical uncertainty, and the lighter orange lines show the total uncertainty about this value.
We find $\tau_{^3P_1} = 21.326 \pm 0.033$~$\mu$s, within the combined uncertainty of the value reported in Ref.~\cite{nicholsonSystematicEvaluationAtomic2015}.
\subsection{Magnetic field and the tensor shift contribution}
During the decay, any static magnetic field or vector and tensor shifts from the lattice light will lead to radiation pattern dynamics, creating a systematic error in the lifetime measurement. 
Applying a $1$~G bias field during the decay, we observe quantum beats in the collected radiation consistent with twice the Larmor frequency of $^3P_1$ $F=9/2$. 
The data used to extract the $^3D_1$ was taken with a magnetic field below $2$~mG and in a $10$~$E_r$ lattice depth.
Nevertheless, field fluctuations and the non-zero lattice depth can lead to systematic bias.
\par We estimate this error by fitting the double exponential model to numerically simulated radiation intensity~\cite{asenjogarciaOpticalWaveguidingAtomic2019,steckQuantumAtomOptics2022} under the influence of such effects. 
The magnetic field, thermal distribution in the lattice, and the excitation to the $^3D_1$ state all contain fluctuations that can greatly affect the resultant radiation. 
We use a Monte Carlo method to sample these fluctuating distributions and capture this effect in the averaged decay. 
\par We consider five hyperfine resolved levels: $^3D_1, F= 11/2$, $^3P_1, F= 11/2$, $^3P_1, F= 9/2$, $^3P_0, F= 9/2$, $^1S_0, F= 9/2$. 
As in the experiment, we start with uniform spin mixture in the $^1S_0$ state then excite the atoms to $^3P_0$ state with equal $\sigma^{\pm}$ polarization. 
The excitation fraction in $^3P_0$  is 0.1 for this simulation. 
The atoms are then excited to $^3D_1$ state with a resonant pulse of various durations to capture the varying excitation fraction. 
\par
\begin{figure}[th!]
    \includegraphics[width=12 cm]{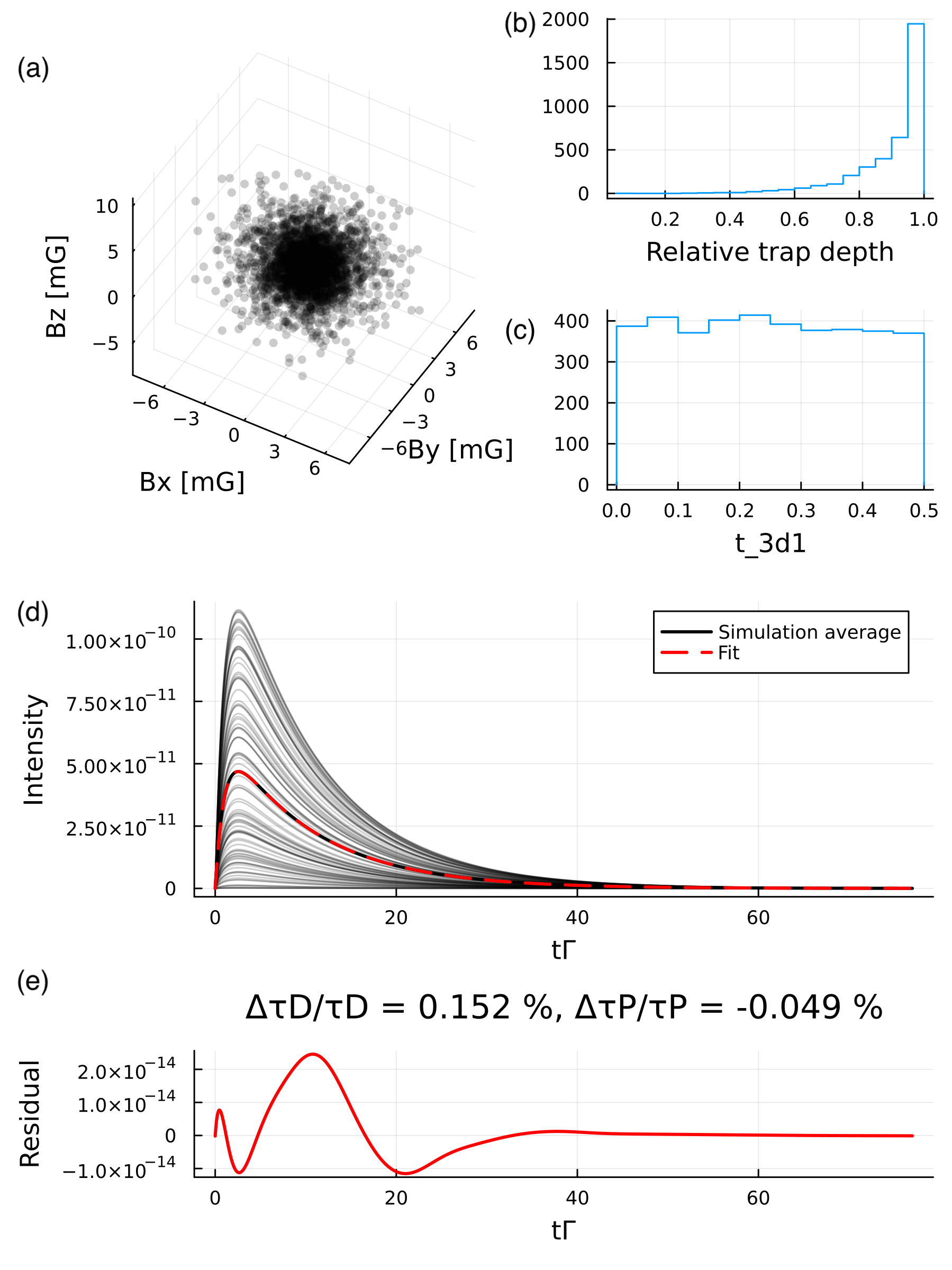}
    \caption{
        Summary of the decay simulation to determine the quantum beat effect. 
        The distribution of sampled systematic error sources are shown in (a-c). 
        (a) is for the magnetic field, (b) is for the lattice depth relative to the maximum lattice depth of 10~\Er, (c) is the pulse duration (0.5 corresponds to the maximum excitation). 
        (d) the gray traces are 100 out of about 4000 simulated radiation decay traces, the black solid line is the average of all simulated decays and the red dashed line is a double exponential fit to this curve. 
        The intensity is in arbitrary units, and $t\Gamma$ is the time normalized to the total decay rate of the $^3D_1$ state. 
        (e) Fit residual of (d) and its fractional error from the true value. For the true value, we use the ratio of the decay rate of $^3D_1$ to the decay rate of $^3P_1$ state, which is 0.1. 
        This value is an arbitrary choice close to the measured value.
        }
    \label{fig:decay_simulation_summary}
\end{figure}

 The simulation result is shown in Figure~\ref{fig:decay_simulation_summary}. 
The detector position, where we calculate the intensity, has a $4^{\circ}$ tilt from the quantization axis (parallel to the lattice polarization), taking into account the geometry of the experiment (the distance from the atom is in far-field regime). 
We estimate the amplitude of the magnetic field noise with a Hall effect sensor. 
For the tensor shift coefficient, we use theoretically calculated polarizability for each state~\cite{Filinprivatecommunication}. 
We extract the lifetime of the $^3D_1$, $\tau D$, by fitting the averaged radiation intensity to the double-exponential model with a least squares method.
The fitted value is compared with the exact value used in the simulation. 
We find fractional error of 0.15 percent, and we treat this value as the systematic error from the magnetic field and the tensor shift.
\subsection{Lifetime Uncertainty Budget}
The total uncertainty in the $^3D_1$ lifetime is dominated by statistics and quantum beats, as shown Table \ref{tab:3d1Budget}.
We consider two other smaller sources of uncertainty, photon counter timing and finite pulse duration.
The multi-channel scaler quotes a $<710$~ps rms timing jitter, which we take as a conservative bound on the timing error. 
As described in Ref.~\cite{nicholsonSystematicEvaluationAtomic2015}, finite pulse duration modifies the decay dynamics from a true double exponential.
To circumvent this, we cut out $500$~ns about this laser pulse.
Residual error from this approach is below $0.1$~ns.
The final $^3D_1$ lifetime is $2.1560 \pm 0.0052$~$\mu$s.
\begin{table}[h]

\begin{tabular}{|c|c|}
\hline
Effect & Uncertainty (ns) \\
\hline \hline

          Statistical &               4.0 \\
        Quantum beats &               3.2 \\
Photon counter timing &               0.7 \\
Finite pulse duration &               0.1 \\ \hline
                Total &               5.2 \\ \hline
\end{tabular}
\caption{\label{tab:3d1Budget}%
$^3D_1$ lifetime uncertainty}
\end{table}
\section{Temperature Measurement}
The two in-vacuum temperature sensors are thin-film platinum resistance thermometers (TFPRTs) from YAGEO-Nexensos (part no. 32208519).
The two chosen sensors were selected from a batch of 10 sensors after observing the measurement stability after numerous thermal cycles up to $200$~$^{\circ}$C.
We use a dry-block technique to calibrate the TFPRTs. 
Two NIST-Gaithersburg calibrated platinum wire wound temperature sensors, labeled R17 and R18 in Ref.~\cite{Tew2015}, are mounted in a temperature controlled copper block. 
These sensors were re-calibrated in 2020, and once mounted in the block agree within their uncertainty.
For calibration, the TFPRTs are inserted between the two wire wound sensors.
The block and environment are varied from 16 to 31~$^{\circ}$C, and the results are fit with the Callendar–Van Dusen equation.
The total calibration uncertainty for both sensors at $20$~$^{\circ}$C is $2$~mK. 
The environment of the dry block is controlled to $200$~mK, leading to a differential immersion error between the three sensors of $1.2$~mK. 
Once installed in-vacuum and baked over a prolonged period at $150$~$^{\circ}$C the sensors continue to agree within their combined uncertainty.
The final temperature we report is an average of both in-vacuum sensors using calibration curves derived from the two wire wound sensors.
The combined uncertainty and mean temperature is determined by a linear pool technique.
\par Using an in-vacuum translation stage, we bound the immersion error by measuring temperature gradients along the translation axis. 
We match the temperature of the fully retracted position with the temperature in the main chamber.
After the 72 hour temperature measurement presented in the main text, the retracted position differed by $3$~mK.
We take this difference as a conservative bound of immersion error.
\par The two temperature sensors on the translation stage have different lines-of-sight to the vacuum chamber and thus are sensitive to thermal gradients. 
We can rotate this probe to modify this differential temperature coupling. 
Within the sensor uncertainties, we have noticed no temperature differences in any orientation, confirming the lack of thermal gradients. 
\par The wire wound sensors are calibrated to the ITS-90 temperature scale, which deviates from true thermodynamic temperature. 
Near room temperature, this correction is quadratic in temperature~\cite{Tew2015}.
At our operational temperature, this is a $2.8$~mK correction with a $0.4$~mK uncertainty.
\par
\begin{figure}[t!]
    \includegraphics[width=10.16 cm]{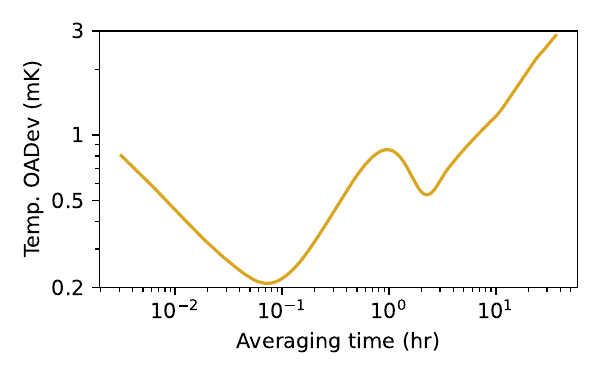}
    \caption{
Overlapping Allan deviations (OADev) of the temperature measured in vacuum over $72$ hours. 
The temperature averages as white noise under 5 minutes.
Periodic temperature fluctuations of the environment contributes to larger instability over the course of an hour.
Temperature drift on the day scale leads to the greatest instability.
For our uncertainty budget, we take the temperature instability at 12 hours.
        }
    \label{fig:temp_oadev}
\end{figure}

 With active chamber temperature stabilization, the long term drift is below $3$~mK$/$day.
Nevertheless, it is critical to periodically measure the temperature at the atoms with the frequency of this measurement determined by the desired accuracy.
Assuming a temperature measurement every $12$ hours, an overlapping Allan deviation of the data presented in the main text estimates this uncertainty at $1.4$~mK.
A complete overlapping Allan deviation is presented in Fig.~\ref{fig:temp_oadev}.
Over short time scales, the temperature averages as white noise. 
At longer time scales, temperature flicker and drift lead to greater instability.
\par The total uncertainty budget is presented in table \ref{tab:tempBudget}.
The operational temperature is $293.2815 \pm 0.0041$~K, with the measurement uncertainty alone account for a BBR fractional shift uncertainty of $2.8 \times 10^{-19}$.

\begin{table}[h]

\begin{tabular}{|c|c|}
\hline
              Uncertainty source &  Uncertainty (mK) \\
\hline \hline
               Calibration error &               2.0 \\
JILA calibration immersion error &               1.2 \\
                 Immersion error &               3.0 \\
               ITS-90 correction &               0.4 \\
   12 hour temperature instability &             1.4 \\ \hline
               Total uncertainty &               4.1 \\
               \hline 
\end{tabular}
\caption{\label{tab:tempBudget}%
Temperature uncertainty sources for the two in-vacuum sensors.}
\end{table}

\section{Dynamic BBR shift}
\subsection{Method}
To determine the dynamic BBR shift, we closely follow the technique presented in Ref.~\cite{lisdat_bbr_2021}. 
The goal of this approach is to estimate relevant dipole matrix elements based upon many experimental observations in a self-consistent manner. 
We treat the important dipole matrix elements as Gaussian variables and compute the experimental observables using the sum-over-states polarizability formula: 
\begin{equation}
    \alpha_i(\nu)=\alpha_i^{\text{core}}+\frac{\epsilon_0 c^3}{(2 \pi)^3} \sum_k \frac{2 J_k+1}{2 J_i+1} \frac{A_{k i}}{\nu_{i k}^2\left(\nu_{i k}^2-\nu^2\right)}.
    \label{eq:polarizability}
\end{equation}
For state $i$, $\alpha_i^{\text{core}}$ is the core polarizability, $A_{k i}$ are the Einstein A coefficients to states $k$, $\nu_{i k}$ are the transition frequencies, and $J_i$ and $J_k$ are the angular momentum of the states. 
The sum is over all the intermediate transitions. 
With this formula we compute various observables $\{y_j'\left(\{A_{k}\}\right)\}$ that depend on $\alpha_i(\nu)$, including the magic wavelength and DC polarizability of $^1S_0$ state in addition to measured A coefficients. 
In this work, we use an updated set of observables $\{y_j\}$ shown in Tab.~\ref{tab:constraints}. 
To implement this procedure, we minimize a $\chi^2$ function,
\begin{equation}
    \chi^2 = \sum_j \left(\frac{y_j'\left(\{A_{k}\}\right) - y_j}{\sigma_j}\right)^2,
    \label{eq:chisq}
\end{equation}
where $y_j$ is the experimental value and $\sigma_j$ is its uncertainty. 
Through this $\chi^2$ minimization, we determine a set of A coefficients that best describes the observed quantities $\{y_j\}$.
\begin{table}[h!]
    \begin{tabular}{ccc}
        \hline \hline
        $y_j$ & Value & Reference\\
        \hline 
        $A[(5s5p)^1P_1 \rightarrow (5s^2)^ 1S_0]$ at 461~nm & $1.9001(14) \times 10^8~s^{-1}$ & \cite{yasudaPhotoassociationSpectroscopySr2006}\\
        $A[(5s5p)^3P_1 \rightarrow (5s^2)^ 1S_0]$ at 689~nm  & $46888(68)~s^{-1}$ & This work\\
        $A[(5s4d)^3D_1 \rightarrow (5s5p)^ 3P_0]$ at 2603~nm  & $2.7619(64)\times10^5~s^{-1}$ & This work\\
        $A[(5s5s)^3S_1 \rightarrow (5s5p)^ 3P_0]$ at 679~nm & $8.348(66) \times 10^6~s^{-1}$ & \cite{heinzStateDependentOpticalLattices2020}\\
        $\alpha ((5s^2)^1S_0, \nu=0)$ & $3.07(24) \times 10^{-39}~\mathrm{C m^{2} V^{-1}}$ &  \cite{schwartzMeasurementStaticElectric1974}\\
        $\Delta \alpha (\nu=0)$ & $4.07873(11) \times 10^{-39}~\mathrm{C m^{2} V^{-1}}$ & \cite{middelmann_static_2012}\\
        Magic wavelength near 813~nm, $\nu^{\text{magic}}_{813}$ & $368~554~825.9(4)~\mathrm{MHz}$ & \cite{kimSr1ACStark}\\
        $\partial\Delta\alpha / \partial \nu (\nu = \nu^{\text{magic}}_{813})$ & $1.859(5)\times 10^{-11}$ & \cite{kimSr1ACStark}\\
        Magic wavelength near 390~nm, $\nu^{\text{magic}}_{390}$& $768917(18)~\mathrm{GHz}$ & \cite{takamotoProspectsOpticalClocks2009}\\
        Tune-out wavelength near 689~nm, $\nu^{\text{to}}$ & $434~972~130(10)~\mathrm{MHz}$ & \cite{heinzStateDependentOpticalLattices2020}\\
        $\alpha({}^3P_0, \nu=\nu^{\text{to}})$& $2.564(13)\times10^{-38}~\mathrm{C m^{2} V^{-1}}$ & \cite{heinzStateDependentOpticalLattices2020}\\
        \hline \hline
    \end{tabular}
    \caption{Experimental observations $\{y_j\}$ used to extract $\{A_j\}$.
    We use the 2603~nm and 689~nm transition matrix elements determined in this work and the light shift measurements from Ref.~\cite{kimSr1ACStark}. \label{tab:constraints}}
\end{table}
\par We use a Monte Carlo sampling technique to estimate the uncertainty of the $\{A_j\}$~\cite{lisdat_bbr_2021,middelmann_static_2012}. 
We randomly draw a set of experimental observations, $\{Y_i\}$, from their distribution (e.g. $Y_i \sim \mathcal{N}(y_i, \sigma_i)$, where $\mathcal{N}(\mu, \sigma)$ is a Gaussian distribution with mean, $\mu$ and standard deviation, $\sigma$), and then minimize the $\chi^2$ function with $\{Y_i\}$ to find the best set of $\{A_j\}$.
We repeat this procedure thousands of times, and then use the resultant distribution to estimate the uncertainty of the $\{A_j\}$.
\par
\begin{figure*}[h]
    \includegraphics[width=\textwidth]{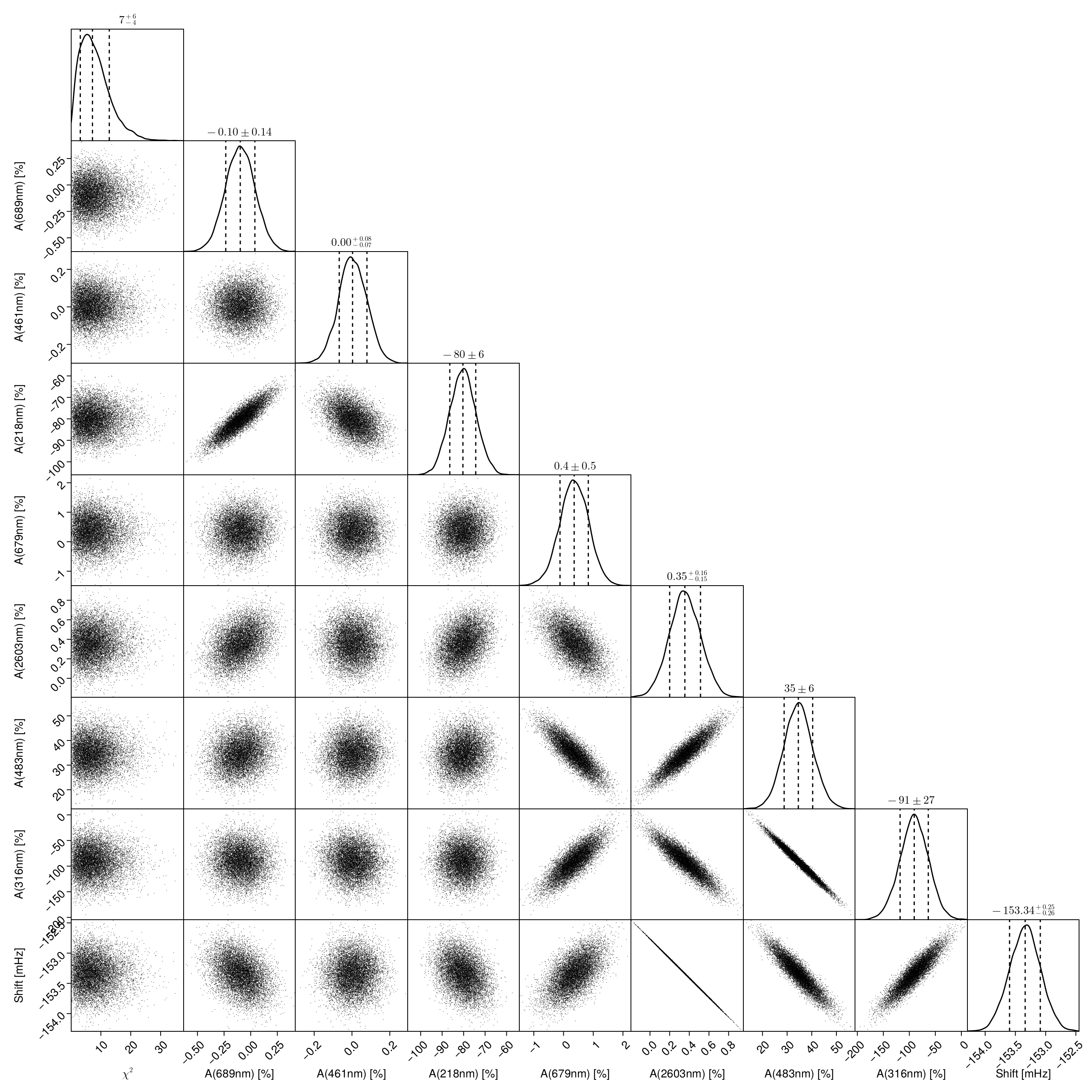}
    \caption{Correlation plot between the fitted $\{A_j\}$.
    In this fit, the magic wavelength measurement near 390~nm is removed. 
    The dots in the cells shows every Monte Carlo sample. 
    The diagonal is the histogram of the matrix element and the titles show (16, 50, 84)\% quantile. 
    The percentage values are the deviation of the fitted value relative to the mean value of the initial condition (see Tab.~\ref{tab:matrix_elements}). 
    A(218~nm), A(483~nm) and A(316~nm) show large deviation because we do not include experimental observations of these coefficients in the fit. 
    $\chi^2$ is calculated using Eq.~\eqref{eq:chisq}.}
    \label{fig:corr_plot}
\end{figure*}

 An example fit result of $\{A_j\}$ is summarized in Tab.~\ref{tab:matrix_elements}. 
In this fitting realization, the magic wavelength measurement near 390~nm is removed. 
This condition is the same as Fit 4 in Ref.~\cite{lisdat_bbr_2021}.
Monte Carlo results are also shown in Fig.~\ref{fig:corr_plot} as a correlation plot. 
As expected, the dynamic shift depends strongly on the $2.6$~$\mu$m A coefficient, with other transitions contributing far less weight.

\begin{table}
    \setlength{\tabcolsep}{8pt}
    \renewcommand{\arraystretch}{0.8}
    \begin{tabular}{ccccccc}
        \hline \hline
        $i$ & $k$ & $D$ [a.u.] & $\Delta D$ [a.u.] & $E_{ki}~\mathrm{[cm^{-1}]}$ & $\lambda$ [nm] & Fit \\
        \hline 
        $(5s^2)^ 1S_0$ & $(5s5p)^ 3P_1$ & $0.1508$ & $0.0001$ & $14504.3380$ & $689.4489$ & true\\
        $(5s^2)^ 1S_0$ & $(5s5p)^ 1P_1$ & $5.2479$ & $0.0019$ & $21698.4520$ & $460.8624$ & true\\
        $(5s^2)^ 1S_0$ & $(5s6p)^ 1P_1$ & $0.2664$ & - & $34098.4040$ & $293.2689$ & -\\
        $(5s^2)^ 1S_0$ & $(5s7p)^ 1P_1$ & $0.3650$ & - & $38906.8580$ & $257.0241$ & -\\
        $(5s^2)^ 1S_0$ & $(5s8p)^ 1P_1$ & $0.5900$ & - & $42462.1360$ & $235.5039$ & -\\
        $(5s^2)^ 1S_0$ & $(5s9p)^ 1P_1$ & $0.4575$ & - & $43328.0400$ & $230.7974$ & -\\
        $(5s^2)^ 1S_0$ & $(5s10p)^ 1P_1$ & $0.3394$ & - & $43938.2010$ & $227.5924$ & -\\
        $(5s^2)^ 1S_0$ & $(5s11p)^ 1P_1$ & $0.2505$ & - & $44366.4200$ & $225.3957$ & -\\
        $(5s^2)^ 1S_0$ & $(5s12p)^ 1P_1$ & $0.1996$ & - & $44675.7370$ & $223.8351$ & -\\
        $(5s^2)^ 1S_0$ & $(5s13p)^ 1P_1$ & $0.1602$ & - & $44903.5000$ & $222.6998$ & -\\
        $(5s^2)^ 1S_0$ & $(5s14p)^ 1P_1$ & $0.1375$ & - & $45075.2900$ & $221.8510$ & -\\
        $(5s^2)^ 1S_0$ & $(5s15p)^ 1P_1$ & $0.1167$ & - & $45207.8300$ & $221.2006$ & -\\
        $(5s^2)^ 1S_0$ & $(4d5p)^ 1P_1$ & $0.6005$ & - & $41172.0540$ & $242.8832$ & -\\
        $(5s^2)^ 1S_0$ & Rydberg \& cont. $^1P_1$ & $0.7037$ & $0.0213$ & $45932.2036$ & $217.7122$ & true\\
        \hline
        $(5s5p)^ 3P_0$ & $(5s6s)^ 3S_1$ & $1.9718$ & $0.0048$ & $29038.7730$ & $679.2894$ & true\\
        $(5s5p)^ 3P_0$ & $(5s7s)^ 3S_1$ & $0.6099$ & - & $37424.6750$ & $432.7661$ & -\\
        $(5s5p)^ 3P_0$ & $(5s8s)^ 3S_1$ & $0.2735$ & - & $40761.3720$ & $378.1595$ & -\\
        $(5s5p)^ 3P_0$ & $(5s9s)^ 3S_1$ & $0.1849$ & - & $42451.1600$ & $355.4462$ & -\\
        $(5s5p)^ 3P_0$ & $(5s10s)^ 3S_1$ & $0.1373$ & - & $43427.4400$ & $343.5254$ & -\\
        $(5s5p)^ 3P_0$ & $(5p^2)^ 3P_1$ & $2.4824$ & - & $35400.1050$ & $474.3248$ & -\\
        $(5s5p)^ 3P_0$ & $(4d^2)^ 3P_1$ & $1.6216$ & - & $44595.9200$ & $330.2683$ & -\\
        $(5s5p)^ 3P_0$ & $(5s4d)^ 3D_1$ & $2.6906$ & $0.0021$ & $18159.0400$ & $2603.1274$ & true\\
        $(5s5p)^ 3P_0$ & $(5s5d)^ 3D_1$ & $2.7249$ & $0.0800$ & $35006.9080$ & $483.3393$ & true\\
        $(5s5p)^ 3P_0$ & $(5s6d)^ 3D_1$ & $1.1388$ & - & $39685.8300$ & $394.1924$ & -\\
        $(5s5p)^ 3P_0$ & $(5s7d)^ 3D_1$ & $0.7537$ & - & $41864.3540$ & $363.0180$ & -\\
        $(5s5p)^ 3P_0$ & $(5s8d)^ 3D_1$ & $0.5475$ & - & $43066.7000$ & $347.8359$ & -\\
        $(5s5p)^ 3P_0$ & $(5s9d)^ 3D_1$ & $0.4238$ & - & $43804.8900$ & $339.1281$ & -\\
        $(5s5p)^ 3P_0$ & Rydberg \& cont. $^3S_1$ & $0.2904$ & - & $45932.2036$ & $316.3086$ & -\\
        $(5s5p)^ 3P_0$ & Rydberg \& cont. $^3D_1 $ & $0.4247$ & $0.0581$ & $45932.2036$ & $316.3086$ & true\\
        \hline \hline
    \end{tabular}
 
    \caption{The reduced dipole matrix elements after the fitting procedure.
    The $D$ is the reduced matrix element in atomic units, $\Delta D$ is its uncertainty from Monte Carlo sampling, $E_k$ is the transition energy, and $\lambda$ is the transition wavelength. 
    The ``Fit" column indicates whether the transition is used as a free parameter in the fit. 
    The values not used in the fit are included in the polarizability calculation as a constant. 
    The constant values are collected and cited in Ref.~\cite{lisdat_bbr_2021}.
   \label{tab:matrix_elements}}
\end{table}

\subsection{Model validation and the final result}
\begin{figure}[h!]
    \includegraphics[width=15cm]{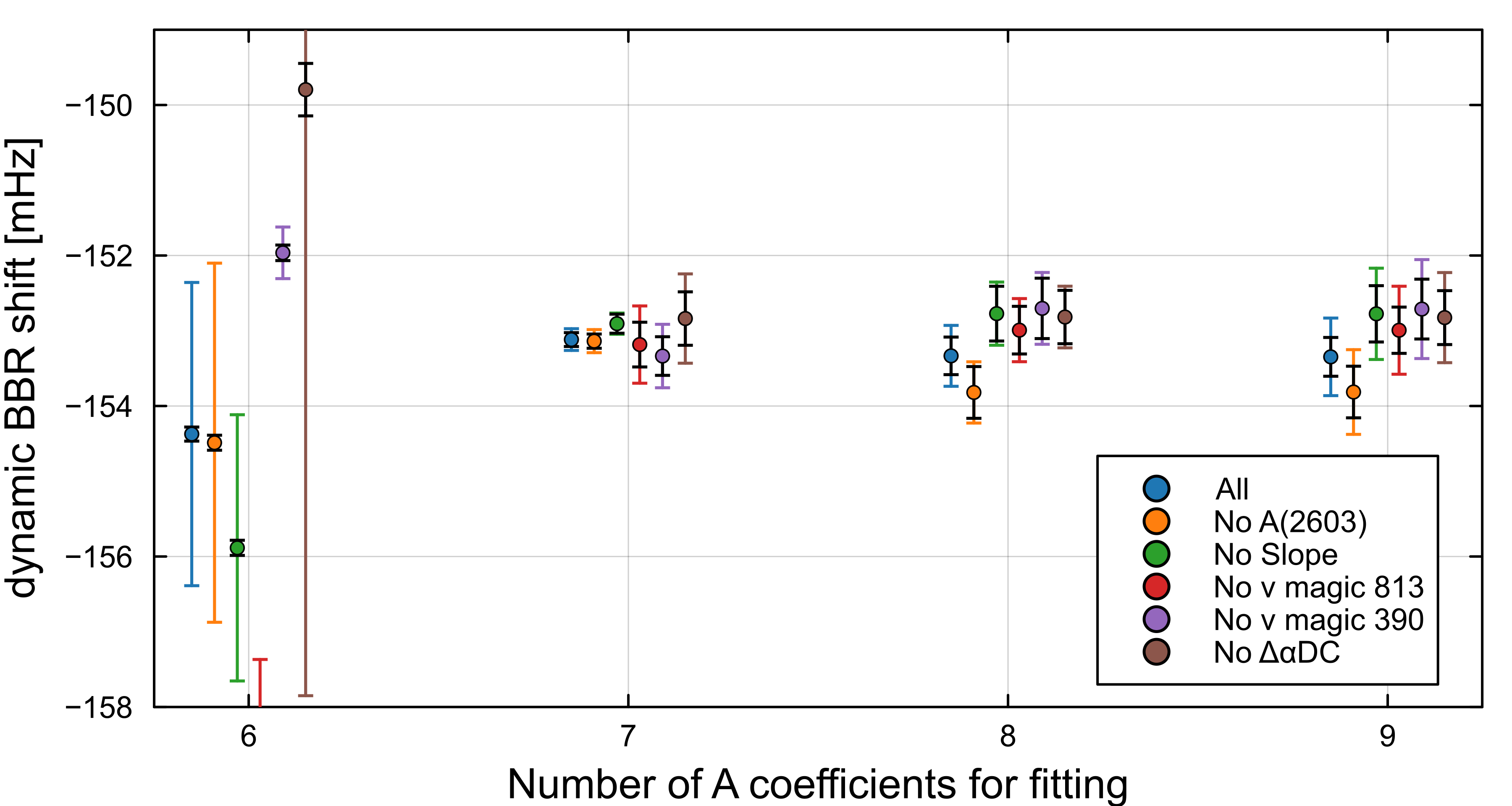}
    \caption{
    Validation of the model and the ``leave-one-out'' approach. 
    The dynamic BBR shift at $300$~K calculated using different sets of $\{A_j\}$ and $\{y_j\}$ is presented. 
    The different choices of $\{y_j\}$ is shown as point colors (horizontally shifted for the visualization, see also Tab.~\ref{tab:constraints}). 
    The black errorbars are the uncertainty estimated from the Monte Carlo propagation and the colored errorbars are inflated by the reduced $\chi^2$.
        }
    \label{fig:cross_validation}
\end{figure}

Equation~\eqref{eq:polarizability} used to construct the fitting model requires an infinite number of transitions, especially to high-lying states. 
For practical purposes, we are limited to a finite number of lines. 
To find the optimal number of transitions, we perform ``leave-one-out" tests to find and also to extract the final dynamic correction. 
For a given set of $\{A_j\}$, we run the fitting routine while removing one data ($\{y_j\}$) at a time. 
The result is presented in Fig.~\ref{fig:cross_validation} and Tab.~\ref{tab:matrix_elements_final}.
Using six or fewer $A_j$ values results in a poor fit with large error and scatter.
As we include more $A$ coefficients, the fit is stabilized.
We decide to use seven $A$ coefficients, where we find the optimal uncertainty with reasonable scatter.
The set of $\{A_j\}$ is the same as Tab.~\ref{tab:matrix_elements}.
We fix $A[(5s5d)^3D_1 \rightarrow (5s5p)^ 3P_0]$ for six-coefficient fits, and free $A[(5s6d)^3D_1 \rightarrow (5s5p)^ 3P_0]$ and $A$[Rydberg \& cont. $^3S_1$] for eight- and nine-coefficient fits.
Because they serve as an effective coefficient for many states, we choose to vary the high-lying states' A coefficients first when increasing the number of coefficients.
The next priority goes to the coefficients with larger weight in the total shift.
\par We choose $A[(5s4d)^3D_1 \rightarrow (5s5p)^ 3P_0]$, $\partial\Delta\alpha / \partial \nu $, $\Delta \alpha (\nu=0)$, $\nu^{\text{magic}}_{813}$, and $\nu^{\text{magic}}_{390}$ for this test, based on the observation that these measurements provide the strongest constraints on the dynamic coefficient. 
$\nu^{\text{magic}}_{390}$ is also included as it likely has the largest sensitivity to the high-lying transitions for all of the experimental observations.
\par In order to account for the scatter and different magnitude of error from each calculation condition, we use the weighted mean as the final dynamic coefficient value and the averaged error (inflated by the square root of the reduced $\chi^2$) as the final error, treating uncertainties as fully correlated Gaussians.
The final value of the dynamic BBR correction at $300$~K is $-153.06(32)$~mHz. 
The mean value deviates considerably from the previous estimate $-150.51(43)$~mHz~\cite{lisdat_bbr_2021}, mainly due to the larger matrix element of $(5s4d) {}^3D_1\rightarrow (5s5p) {}^3P_0$ transition revised in this work.
\par In future work, it will be interesting to include theoretically calculated values for high lying state $A$ coefficients in this comprehensive fit. 
The same fitting procedure as outlined above can be applied, using a mixture of experimental and theory values. 
Of course, this approach requires caution regarding the treatment of experimental and theoretical errors.
Nevertheless, it will provide useful insights for future measurements to the properties of this important atom.  

\begin{table*}
    \scriptsize
    \begin{tabular}{ccccccc}
        \hline \hline
        $k$ & All $\{y_j\}$ & No $A_{2603~nm}$ & No $\partial\alpha/\partial\nu$ & No $\nu^{\mathrm{magic}}_{813}$ & No $\nu^{\mathrm{magic}}_{390}$ & No $\Delta \alpha (0)$\\
        \hline
        $(5s5p)^ 3P_1$ & 0.15075(0.0001) & 0.15077(0.00011) & 0.15083(0.00011) & 0.15076(0.0001) & 0.15077(0.0001) & 0.15077(0.0001)\\
        $(5s5p)^ 1P_1$ & 5.2478(0.0019) & 5.2478(0.0019) & 5.2477(0.0019) & 5.2479(0.0019) & 5.2479(0.0019) & 5.2478(0.0019)\\
        $(5s6p)^ 1P_1$ & 0.2664 & 0.2664 & 0.2664 & 0.2664 & 0.2664 & 0.2664\\
        $(5s7p)^ 1P_1$ & 0.365 & 0.365 & 0.365 & 0.365 & 0.365 & 0.365\\
        $(5s8p)^ 1P_1$ & 0.59 & 0.59 & 0.59 & 0.59 & 0.59 & 0.59\\
        $(5s9p)^ 1P_1$ & 0.4575 & 0.4575 & 0.4575 & 0.4575 & 0.4575 & 0.4575\\
        $(5s10p)^ 1P_1$ & 0.3394 & 0.3394 & 0.3394 & 0.3394 & 0.3394 & 0.3394\\
        $(5s11p)^ 1P_1$ & 0.2505 & 0.2505 & 0.2505 & 0.2505 & 0.2505 & 0.2505\\
        $(5s12p)^ 1P_1$ & 0.1996 & 0.1996 & 0.1996 & 0.1996 & 0.1996 & 0.1996\\
        $(5s13p)^ 1P_1$ & 0.1602 & 0.1602 & 0.1602 & 0.1602 & 0.1602 & 0.1602\\
        $(5s14p)^ 1P_1$ & 0.1375 & 0.1375 & 0.1375 & 0.1375 & 0.1375 & 0.1375\\
        $(5s15p)^ 1P_1$ & 0.1167 & 0.1167 & 0.1167 & 0.1167 & 0.1167 & 0.1167\\
        $(4d5p)^ 1P_1$ & 0.6005 & 0.6005 & 0.6005 & 0.6005 & 0.6005 & 0.6005\\
        Rydberg \& cont. $^1P_1$ & 0.693(0.021) & 0.71(0.022) & 0.766(0.024) & 0.694(0.021) & 0.704(0.021) & 0.707(0.022)\\
        \hline
        $(5s6s)^ 3S_1$ & 1.9755(0.0027) & 1.9762(0.0028) & 1.9668(0.0045) & 1.9751(0.0034) & 1.9718(0.0048) & 1.9763(0.0029)\\
        $(5s7s)^ 3S_1$ & 0.6099 & 0.6099 & 0.6099 & 0.6099 & 0.6099 & 0.6099\\
        $(5s8s)^ 3S_1$ & 0.2735 & 0.2735 & 0.2735 & 0.2735 & 0.2735 & 0.2735\\
        $(5s9s)^ 3S_1$ & 0.1849 & 0.1849 & 0.1849 & 0.1849 & 0.1849 & 0.1849\\
        $(5s10s)^ 3S_1$ & 0.1373 & 0.1373 & 0.1373 & 0.1373 & 0.1373 & 0.1373\\
        $(5p^2)^ 3P_1$ & 2.4824 & 2.4824 & 2.4824 & 2.4824 & 2.4824 & 2.4824\\
        $(4d^2)^ 3P_1$ & 1.6216 & 1.6216 & 1.6216 & 1.6216 & 1.6216 & 1.6216\\
        $(5s4d)^ 3D_1$ & 2.6887(0.0008) & 2.68891(0.00082) & 2.6869(0.0011) & 2.6893(0.0027) & 2.6906(0.0021) & 2.6863(0.0031)\\
        $(5s5d)^ 3D_1$ & 2.668(0.0079) & 2.6675(0.0078) & 2.683(0.011) & 2.667(0.012) & 2.725(0.08) & 2.6665(0.0081)\\
        $(5s6d)^ 3D_1$ & 1.1388 & 1.1388 & 1.1388 & 1.1388 & 1.1388 & 1.1388\\
        $(5s7d)^ 3D_1$ & 0.7537 & 0.7537 & 0.7537 & 0.7537 & 0.7537 & 0.7537\\
        $(5s8d)^ 3D_1$ & 0.5475 & 0.5475 & 0.5475 & 0.5475 & 0.5475 & 0.5475\\
        $(5s9d)^ 3D_1$ & 0.4238 & 0.4238 & 0.4238 & 0.4238 & 0.4238 & 0.4238\\
        Rydberg \& cont. $^3S_1$ & 0.2904 & 0.2904 & 0.2904 & 0.2904 & 0.2904 & 0.2904\\
        Rydberg \& cont. $^3D_1 $ & 0.83(0.016) & 0.834(0.016) & 0.892(0.02) & 0.824(0.017) & 0.425(0.058) & 0.829(0.016)\\
        \hline
        $\nu_{dyn}~300$~K [mHz] & -153.12(0.17) & -153.14(0.17) & -152.91(0.15) & -153.17(0.54) & -153.34(0.42) & -152.84(0.59)\\
        \hline \hline
    \end{tabular}
    \caption{\label{tab:matrix_elements_final} Reduced dipole matrix elements in atomic units from the validation test in Fig.~\ref{fig:cross_validation}. We shows number of A coefficients of seven case, which we use to compute the final BBR shift. The matrix elements without the error are used as the fitting parameter. Other coefficients are fixed as constants.}
\end{table*}


\section{First Order Zeeman Shift}
\begin{figure}[th!]
    \includegraphics[width=8 cm]{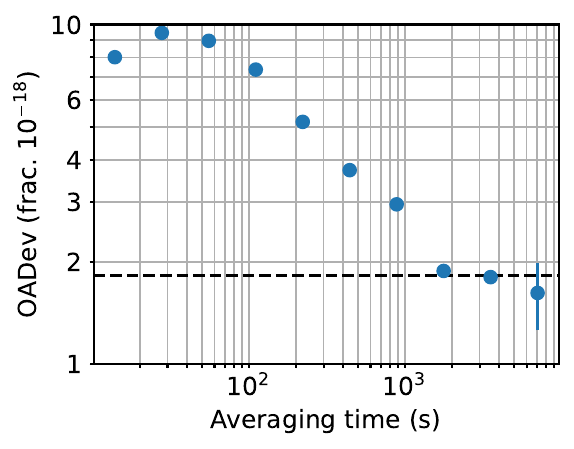}
    \caption{
    Overlapping Allan deviation (OADev) of the frequency difference between subsequent measurements of the $|^1S_0 ~ m_F = - 5/2 \rangle \rightarrow | ^3P_0 ~ m_F = - 7/2 \rangle$ transition.
    The dashed line represents a flicker floor of this measurement at $1.8 \times 10^{-18}$.
    We treat this flicker floor as the first order Zeeman shift uncertainty.}
    \label{fig:1st_order_zeeman}
\end{figure}

While we use the least magnetically sensitive Sr clock transition, drifting magnetic fields and low frequency noise may lead to an uncompensated first order Zeeman shift~\cite{oelkerDemonstration10172019}.
Our typical clock measurement sequence alternates between  $|^1S_0 ~ m_F = + 5/2 \rangle \rightarrow | ^3P_0 ~ m_F = + 3/2 \rangle$ and $|^1S_0 ~ m_F = - 5/2 \rangle \rightarrow | ^3P_0 ~ m_F = - 3/2 \rangle$ transitions.
To characterize the magnetic field environment, we use the 26 times more magnetically sensitive $|^1S_0 ~ m_F = - 5/2 \rangle \rightarrow | ^3P_0 ~ m_F = - 7/2 \rangle$ transition.
This allows us to use the same sample preparation technique as in our standard operation, maintaining an identical duty cycle.
We look at the frequency difference between successive measurements of the same transition, which contains the first order Zeeman coupling at our duty cycle.
An overlapping Allan deviation of this frequency series is shown in Fig.~\ref{fig:1st_order_zeeman}.
The average difference is consistent with $0$, and the measured noise floor of $1.8 \times 10^{-18}$ represents an upper bound on the magnetic field noise. 
After scaling to the operational transition, we find a $7\times10^{-20}$ uncertainty on the first order Zeeman shift.
\section{Background gas shift}
\begin{figure}[h!]
    \includegraphics[width=10.16 cm]{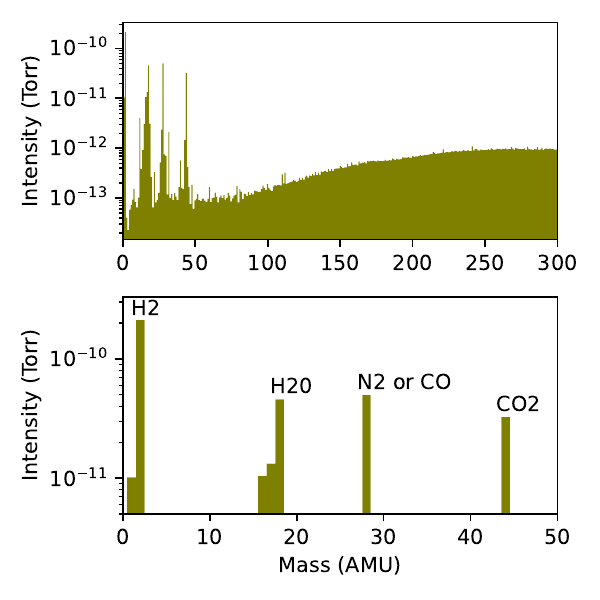}
    \caption{
    Background gas composition of the vacuum system.
    Using a residual gas analyzer equipped with an electron multiplying stage, we measure partial pressures of vacuum contaminants up to 300 atomic mass units (AMU), as shown in the top plot.
    The lower plot highlights the predominant species in the vacuum.
        }
    \label{fig:rga_scan}
\end{figure}

We use a residual gas analyzer (RGA) with an electron multiplier stage to measure the partial pressures of the vacuum up to 300 atomic mass units.
The RGA does not have direct line of sight to the atoms, so the absolute pressure values it reads are likely inaccurate.
Mounted in an auxiliary chamber that also holds the temperature sensor and translation arm, we believe the local pressure and non-hydrogen vacuum contaminants the RGA measures to be significantly higher than in the primary chamber. 
The partial pressures for each atomic mass are shown in Fig. \ref{fig:rga_scan}.
We confirm that the dominant species in the vacuum is hydrogen, with a partial pressure of $2 \times 10^{-10}$ Torr at the RGA.
Other significant vacuum contaminants include H$_2$O, N$_2$, CO, and CO$_2$, all with partial pressures below $5 \times 10^{-11}$ Torr.
The background gas coefficient reported in~\cite{alves_bgshift} was measured in a hydrogen dominated system, so it is valid in this environment.
\par
\begin{figure}[h!]
    \includegraphics[width=10.16 cm]{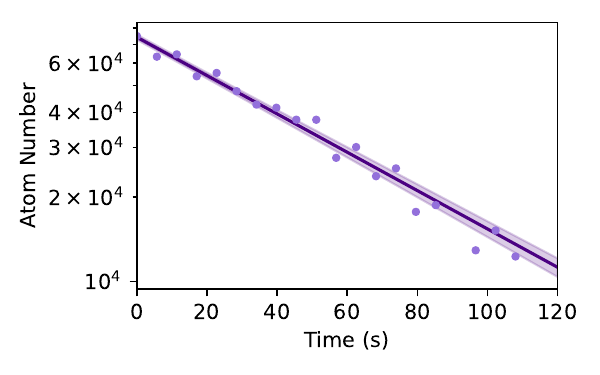}
    \caption{
Vacuum lifetime of trapped atoms. 
The solid line shows an exponential fit to the atom number, and the shaded area shows the fit uncertainty.
The vacuum lifetime is $63.6 \pm 2.5$~s.
        }
    \label{fig:vacuum_lifetime}
\end{figure}

 We measure the vacuum lifetime by measuring the atom number as a function of hold time in a deep lattice.
Background gas collisions with this trapped sample lead to atom loss over time.
We fit this data to an exponential decay and find a vacuum lifetime of $63.6 \pm 2.5$~s.
The results are presented in Fig.~\ref{fig:vacuum_lifetime}. 
This measurement is vacuum limited, as slightly increasing pressure leads to a shorter measured lifetime.
With this lifetime, the background gas shift is $(-4.7\pm0.5) \times 10^{-19}$.


\clearpage
\bibliography{bibliography}